\documentclass[preprint,proceedings]{rmaa}

\usepackage{paralist}

\SetYear{2010}
\SetConfTitle{Highlights of Bol. Obs. Tonantzintla y Tacubaya}

\title{"Chemical abundances in \ion{H}{2} regions and their implications" 
{\it Retrospective on: M. Peimbert \& R. Costero,  Boletin de los Observatorios 
de Tonantzintla y Tacubaya\/, Vol. \underbar{5}, 3, 1969}

}

\author{
  M. Peimbert\altaffilmark{1} and A. Peimbert\altaffilmark{1}, 
  }

\altaffiltext{1}{Instituto de Astronom\'\i{}a, U.N.A.M.
  M\'exico, D.F., M\'exico.}

\shortauthor{Peimbert \& Peimbert}

\shorttitle{abundances in \ion{H}{2} regions}

\fulladdresses{
\item M. Peimbert:  
  Universidad Nacional Aut\'o\-no\-ma de M\'exico, 
  Apartado Postal 70-264, 04510
  M\'exico, D.F., M\'exico. (\email{peimbert@astroscu.unam.mx})
\item A. Peimbert:  
  Universidad Nacional Aut\'o\-no\-ma de M\'exico, 
  Apartado Postal 70-264, 04510
  M\'exico, D.F., M\'exico. (\email{antonio@astroscu.unam.mx})}

\listofauthors{M. Peimbert \& A. Peimbert}

\indexauthor{Peimbert, M.}
\indexauthor{Peimbert, A.}

\abstract{We present a review about the relevance of the paper by Peimbert
and Costero (1969), on the chemical abundance determinations of \ion{H}{2} regions.
We analize the observational evidence in favor of the presence of temperature variations
inside gaseous nebulae. We make a brief mention of the methods used to estimate the contribution
of the unobserved ions to the total chemical abundances.}

\resumen{Presentamos una revisi\'on sobre la relevancia del art\'{\i}culo
de Peimbert y Costero (1969), en la determinaci\'on de la
composici\'on qu\'{\i}mica de regiones \ion{H}{2}. Analizamos las evidencias observacionales
en favor de la presencia de variaciones de temperatura en nebulosas gaseosas. Hacemos una
breve menci\'on a los m\'etodos para tomar en cuenta a los iones no observados en la 
determinaci\'on de las abundancias de los elementos. }

\addkeyword{galaxies: abundances}
\addkeyword{galaxies: evolution}
\addkeyword{H II regions}
\addkeyword{H II regions: individual: Orion nebula, M8, M17}
\addkeyword{ISM: abundances}
\addkeyword{Sun: abundances}

\begin{document}
\maketitle

\section{Introduction}
\label{sec:intro}

The paper by M. Peimbert and R. Costero (1969, hereafter PC) presents the determination
of the chemical composition of three galactic \ion{H}{2} regions, the Orion nebula, M8 and M17.
To derive the chemical abundances PC took into account the
presence of temperature variations inside gaseous nebulae based on the definitions
of the average temperature, $T_0$, and the mean square temperature variation, $t^2$,
introduced by Peimbert (1967), that are given by

\begin{equation}
T_0 (V, i) = \frac{\int T_e N_e N_i dV}
{\int N_e N_i dV},
\end{equation}
and
\begin{equation}
t^2 = \frac{\int (T_e - T_0)^2 N_e N_i dV}{T_0^2 \int N_e N_i dV},
\end{equation}
respectively, where $T_e$, $N_e$, and $N_i$ are the electron temperature, the electron density, and the ion density
of the volume element, and $V$ is the observed volume.

PC includes simple equations to determine the chemical abundances in the presence of
temperature variations, additional and more detailed equations to determine the chemical 
abundances have been presented by Peimbert et al. (2002), Ruiz et al. (2003) and Peimbert et al. (2004).

Accurate abundances are needed to test models of: gaseous nebulae, stellar evolution,
galactic chemical evolution, and the observable universe as a whole. The presence of 
temperature variations produces biases in the abundance determinations
derived from collisionally excited lines that are typically in the 0.1 to 0.5 dex range.
PC provided significant advances in the procedures to determine abundances 
of \ion{H}{2} regions and planetary nebulae.

We will mention the observational evidence in favor of large temperature variations based on seven independent
methods, and some of the most relevant results that abundances, derived including the effect of temperature variations, 
have had in different areas of astronomical research. We will define those gaseous nebulae with $t^2$ values $< 0.01$  as small temperature variation objects, and those with $t^2$ values $> 0.01$ as large temperature variation objects.

Most papers dedicated to the determination of abundances in gaseous nebulae assume that temperature variations are
small, and consequently that the abundances can be derived under the assumption of constant temperature. In section 2
we present evidence in favor of large temperature variations inside most
gaseous nebulae. In section 3 we present abundance determinations that agree with the presence of
large temperature variations based on: the comparison of stellar abundances with gaseous nebular abundances,
the comparison of solar and \ion{H}{2} region abundances
within the framework of chemical abundance models of the Galaxy, and the
comparison of the primordial helium abundance derived from \ion{H}{2} regions compared with that derived from the Standard
Big Bang Nucleosynthesis. In section 4 we discuss briefly the ionization correction factor method to derive total
chemical abundances presented in PC.

\section{Simultaneous determinations of $T_0$ and {\lowercase{$t^2$}}}
\label{sec:Temp}

To determine $T_0$ and $t^2$ we need to combine electron temperatures, $T_e$, based on two different
methods: one that weighs preferentially the high temperature regions and one that weighs
preferentially the low temperature regions. Photoionization models of chemically 
homogeneous \ion{H}{2} regions predict $t^2$ values typically in the 0.003 to 0.01 range, 
while observations yield $t^2$ values typically in the 0.02 to 0.06 range. 

These results imply that there are additional heating 
and cooling processes in \ion{H}{2} regions that are not yet included in photoionization models.
Many possible causes of temperature variations have been discussed in the
literature some of them are: X-ray heating, cosmic-ray heating, density variations, deposition of mechanical energy,
deposition of magnetic energy, presence of shadowed regions,
chemical inhomogeneities, dust heating, and transient effects due
to changes in the ionizing flux. The source of large temperature variations in gaseous nebulae
is still an open problem, and the relative importance of the effects causing the variations might be different for each nebula.

In what follows we will mention some relevant papers that, based on seven independent methods, indicate 
the presence of large temperature variations in gaseous nebulae.

\subsection{Temperatures derived from the Balmer continuum and Balmer line intensities together with 
temperatures derived from collisionally excited line intensities}

The large temperature variations idea was based mainly on three results: a) the smaller $T$(Bac) values than 
those derrived from $T$([O III]) and $T$([N II]) (PC and Peimbert 1967) and the $t^2$ values around 0.04 derived from the photoionization models by Hjellming (1966), that included only N, O and Ne as potential coolants of the model \ion{H}{2} regions. In the seventies and eighties most observers obtained $t^2$ values from $T$(Bac) and $T$(O III) of about
0.02 $\pm~0.04$ that were consistent with $t^2$ = 0.00. Moreover photoionization models that 
included more chemical elements than those considered by Hjellming predicted $t^2$ values of about 0.005.
Probably these two results together with the simpler equations to determine chemical abundances led
most workers to adopt $t^2$ = 0.00 and the temperature provided by $T$(O III) to determine the abundances of 
gaseous nebulae.

The accuracy of the temperature determinations improved and Liu and Danziger (1993) by combining $T$(Bac) with $T$(O III) determinations found large $t^2$ values for 14 planetary nebulae with an average value around 0.03 and errors considerably smaller than the $t^2$ values. Similarly Garc\'{i}a-Rojas, Esteban and collaborators (Garc\'{i}a-Rojas 
et al. 2004, 2005, 2006, 2007, Esteban et al. 2004) by combining $T$(Bac) with $T$(O III) and $T$(O II) 
determinations for eight galactic \ion{H}{2} regions found $t^2$ values in the 0.01 to 0.056 range with 
an average value of 0.034, again with errors smaller than the $t^2$ values. Moreover Gonz\'alez Delgado et al. (1994)
studied the giant extragalactic \ion{H}{2} region NGC 2363 and based on measurements of the Paschen discontinuity
found $t^2$ values of 0.064 and 0.098 for knots A and B respectively.

\subsection{Abundances derived from collisionally excited C III lines together with abundances derived from C II
recombination lines}

The $N$(C$^{++}$/$N$(H$^+$) values derived for \ion{H}{2} regions and planetary nebulae 
based on the C~II $\lambda$ 4267 to H$\beta$ intensity
ratio are, in general, higher than those derived from the C III $\lambda\lambda$~
1906 + 1909 to H$\beta$ intensity ratio; in some cases the difference 
reaches a factor of ten. General discussions of this problem have been 
given in the literature (e. g. Torres-Peimbert, Peimbert \& Daltabuit
1980; Kaler 1986; Rola \& Stasinska 1994, Peimbert, Torres-Peimbert, \& Luridiana 1995,
Liu 2006, Peimbert \& Peimbert 2006, and references therein). Several
ideas have been advanced to explain the discrepancy: a) errors in the
atomic data, b) errors in the observations and c) the presence of spatial
temperature variations. 

\begin{table*}[!t]\centering
\setlength{\tabnotewidth}{0.5\columnwidth}
  \tablecols{6}
\setlength{\tabcolsep}{0.5\tabcolsep}
\caption{{\lowercase{$t^2$}} values derived from different methods }
\label{tta:disk}
\begin{tabular}{lr@{$\pm$}lr@{$\pm$}lr@{$\pm$}lr@{$\pm$}l}
    \toprule
{Method} &
\multicolumn{2}{c}{Orion Nebula}  &
\multicolumn{2}{c}{30 Doradus} &
\multicolumn{2}{c}{NGC 5315} &
\multicolumn{2}{c}{NGC 6543}
\\
\midrule
$T$(Bac) and $T_e$([O II],[O III])       & 0.018  & 0.018      & 0.022 & 0.007          & 0.039 & 0.022 & 0.028 & 0.009 \\
$T$(He II) and $T_e$([O II],[O III])     & 0.022  & 0.002  &\multicolumn{2}{c}{\nodata} & 0.060 & 0.007 & 0.035 & 0.014\\
$N$(C$^{++}$)RL and $N$(C$^{++}$)CEL      & 0.039  & 0.011     & 0.056 & 0.040        & 0.063 & 0.035 & 0.036 & 0.010\\
$N$(O$^+$)RL and $N$(O$^+$)CEL & 0.052  & 0.029 & 0.075 & 0.040 &\multicolumn{2}{c}{\nodata} &\multicolumn{2}{c}{\nodata} \\
$N$(O$^{++}$)RL and $N$(O$^{++}$)CEL    & 0.020  & 0.002     & 0.038 & 0.005         & 0.048   &0.004  & 0.024 & 0.008    \\
$N$(Ne$^{++}$)RL and $N$(Ne$^{++}$)CEL  & 0.032  & 0.014  & \multicolumn{2}{c}{\nodata}& 0.068  &0.020 & 0.022 & 0.010   \\
Average                             & 0.022  & 0.002     & 0.033 & 0.005           & 0.051  &0.004 & 0.028 & 0.005\\

\bottomrule
\end{tabular}
\end{table*}

\subsection{Abundances derived from collisionally excited [O III] lines together with abundances derived from O II
recombination lines}

Peimbert, Storey, \& Torres-Peimbert (1993) were the first 
to determine O/H values for gaseous nebulae based on the recombination
coefficients for \ion{O}{2} lines computed by Storey (1994). The temperature dependence
of the \ion{O}{2} lines is relatively weak and very similar to that of the
\ion{H}{1} lines, therefore the O$^{++}$/H$^+$ ratios are independent of the
electron temperature. Alternatively the O$^{++}$/H$^+$ ratios derived
from collisionally excited lines do depend strongly on the average
temperature, $T_0$, and the mean temperature square, $t^2$ (e. g.:
Peimbert 1967, PC, Ruiz et al. 2003, Peimbert et al. 2004).

Garc\'{i}a-Rojas, Esteban and collaborators
(Garc\'{i}a-Rojas et al. 2004, 2005, 2006, 2007, Esteban et al. 2004) observed 8 galactic \ion{H}{2} regions 
and by combining the [O III] lines with the O II lines find $t^2$ values in the 0.020 to 0.046 range
with an average value of 0.038. Similarly, Esteban et al. (2002),  Peimbert (2003),
Bresolin (2007), and Esteban et al. (2009) for 11 extragalactic \ion{H}{2} regions
find $t^2$ values in the 0.027 to 0.124 range with an average value of 0.048.

\subsection{He I recombination lines} 

From a large set of He I lines it is possible to determine $T$(He I). This temperature combined with
$T$([O III]) and $T$([O II])  yields  $T_0$ and $t^2$  (Peimbert, Peimbert, \& Ruiz
2000, Peimbert, Peimbert, \& Luridiana 2002). This method has been applied by Garc\'{i}a-Rojas, Esteban
and collaborators (Garc\'{i}a-Rojas et al. 2004, 2005, 2006, 2007, Esteban et al. 2004) to 7 galactic 
\ion{H}{2} regions yielding $t^2$ values in the 0.017 to 0.046 range, with an average value of 0.027.

This method has also been applied by  Peimbert, Luridiana, and Peimbert (2007a) to five metal poor
extragalactic \ion{H}{2} regions obtaining an average $t^2$ value of 0.026; and by Esteban et al. (2009) to 14 giant extragalactic \ion{H}{2} regions obtaining $t^2$ values in the 0.022 to 0.125 range with an average value of 0.058.

\subsection{Comparison of the $t^2$ values derived from different methods}

In Table 1 we present $t^2$ values determined from six different methods for four well observed objects, two
\ion{H}{2} regions and two planetary nebulae (Esteban et al. 2004, Peimbert 2003, Peimbert et al. 2004,
Georgiev et al. 2008). For each object the different methods yield $t^2$ values that are in very good agrement.
 Moreover the adopted average values show errors
that are from five to twelve times smaller than the $t^2$ determinations. These two results imply that the temperature variations are real and very large.

To explain the presence of large spatial temperature 
fluctuations, considerably higher than those predicted by chemically
homogeneous photoionized models, several possibilities have been
proposed. Reviews on this problem have been presented by many authors see
for example:  Torres-Peimbert, Peimbert, \& Pe\~na (1990), Peimbert, Sarmiento, 
\& Fierro (1991), Esteban (2002), Torres-Peimbert and Peimbert (2003), 
Liu (2006), Peimbert, \& Peimbert (2006).

\subsection{High spatial resolution map of the columnar temperature in the Orion nebula}

O'Dell, Peimbert, and Peimbert (2003), based on observations with the Hubble Space Telescope, presented a high spatial resolution map of the columnar electron temperature
$T_c$ of a region to the south west of the Trapezium in the Orion
Nebula . {From} their $T_c$(O$^{++}$) map of the Orion Nebula, that includes
1.5 x 10$^{6}$ independent temperature determinations, they found that the observed mean square temperature variation in the plane of the sky, $t^2{_a}$ (O$^{++}$), amounts to $ 0.008$.

Based on their $t^2{_a}$ (O$^{++}$) value, together with geometrical considerations and other
observations in the literature, they estimated that  $t^2$(O$^{++}$)$=
0.021$. Note that the total  $t^2$ (O$^{++}$) is larger
than $t^2{_a}$~(O$^{++}$) because in addition to the variations across the
plane of the sky it includes the temperature
variations along the line of sight. {From} their $t^2$ (O$^{++}$) value and comparisons between the
temperatures in the low- and high-ionization zones, the O$^{+}$ and O$^{++}$
zones, they found that $t^2$ (H~{\sc ii})$ = 0.028 \pm 0.006$. Their derived  $t^2$ (H~{\sc ii}) value is 7 times higher than those obtained from homogeneous one-dimensional photoionization models of
the Orion Nebula (e. g. Baldwin et al. 1991, Peimbert et al. 1993).

\subsection{Comparison of the different ways to calibrate Pagel's method}

The most popular metallicity indicator to determine the O/H ratio in extragalactic \ion{H}{2} regions was introduced by Pagel et al.
(1979, see also Edmunds \& Pagel 1984) and is indistinctly known as Pagel's, or $R_{23}$, or $O_{23}$
indicator, where $O_{23}\equiv 
I([$\ion{O}{2}$]\lambda 3727 + [$\ion{O}{3}$]\lambda\lambda 4959, 5007)/ I({\rm H}\beta)$. 
The $O_{23}$ indicator has been calibrated
with the O/H values based on three different methods: a) by using photoionization
models, that we will call $PIM$ calibrations or $PIM$ method, b)
by using observational determinations of the O/H abundances 
based on the electron temperature derived
from the $I(4363)/I(5007)$ [$\ion{O}{3}$] ratio together with  the
$I(3727)/I({\rm H}\beta)$ and the $I(5007)/I({\rm H}\beta)$ line ratios, the so called 
$T(4363)$ method, and c) by using observational determinations of the O/H abundances  
based on the intensity ratio of \ion{O}{2} 
recombination lines to  \ion{H}{1}  recombination lines that has been called the \ion{O}{2}$_{RL}$ method
by Peimbert et al. (2007b). 

These three methods have been discussed by several authors, (e. g. McGaugh 1991, Pilyugin \& Thuan 2005,
Peimbert et al. 2007b, Kewley \& Ellison 2009, and references therein). These three methods produce 
very different O/H calibrations, in extreme cases the differences in the
inferred O/H abundances among these calibrations reach values of 0.7 dex.

Peimbert et al. (2007b) reach the following
conclusions: a) the \ion{O}{2}$_{RL}$ method supports the suggestion that the controversy
produced by the relatively high O/H values predicted
by the $PIM$ calibrations and the relatively low O/H
values predicted by the  $T(4363)$
calibrations are mainly due to temperature variations; b)
the best way to calibrate the $O_{23}$ indicator is to use the
\ion{O}{2}$_{RL}$ method to obtain the O/H values because it is
independent of the temperature structure of the \ion{H}{2} regions; c) the use 
of $T$(4363) values to derive O/H, under the assumption of
constant temperature, provides a lower limit to the O/H abundance ratios; d)
since the nebular lines are less sensitive to $T_0$ and $t^2$ than the auroral lines, 
the model calibrations that adjust the nebular lines are closer 
to the \ion{O}{2}$_{RL}$ calibration than those derived using the 
observed $T$(4363) values; and e) the presence of 
temperature variations affects strongly the $T$(4363) method, weakly
the $PIM$ method, and leave the \ion{O}{2}$_{RL}$
method unaffected, or in other words the \ion{O}{2}$_{RL}$
method is independent of the temperature structure
of the nebula.

\section{Additional support in favor of large temperature variations}
\label{sec:Supp}

\subsection{Comparison of the abundances of the nebula of NGC 6543 with those of its central star}

Table 2 presents the stellar abundances of the planetary nebula NGC 6543 together with the nebular
envelope abundances derived from recombination lines (RC) that do not depend on the temperature structure of the
nebula, and nebular abundances derived from collisionally excited lines (CL) under the assumption that $t^2$ = 0.00
determined by Georgiev et al. (2008).

The  stellar abundances of He, C and O are in excellent agreement with the nebular abundances derived from recombination lines. In particular the similar He/H values imply that there are no He rich inclusions present in the nebula.
On the other hand the C, O and S abundances derived from nebular collisionally excited lines
(under the assumption of $t^2$ = 0.00) are considerably smaller than the stellar abundances. Table 1 
shows the $t^2$ values needed to reach agreement between the recombination and the collisional abundance determinations. These results imply that indeed large temperature variations are present NGC 6543 and that the nebula of
this object is chemically homogeneous.

In addition a $t^2$ value of about 0.028 permits to reach agreement between the nebular 
determination of S and the stellar one. Similarly a $t^2$ value of about 0.028 also permits
to reach agreement between the Ne abundances derived from recombination and collisionally excited lines.

\begin{table}[!t]\centering
\setlength{\tabnotewidth}{0.5\columnwidth}
  \tablecols{6}
\setlength{\tabcolsep}{0.5\tabcolsep}
\caption{Stellar and nebular abundances for NGC 6543\tablenotemark{a} }
\label{tta:disk}
\begin{tabular}{lr@{$\pm$}lr@{$\pm$}lr@{$\pm$}l}
    \toprule
{Element} & 
\multicolumn{2}{c}{Stellar}  &
\multicolumn{2}{c}{Nebular(RC)} &
\multicolumn{2}{c}{Nebular(CL)}
\\
\midrule
He &11.00 &0.04 &11.05 &0.01 &\multicolumn{2}{c}{\nodata}\\
C  & 9.03 &0.10 & 8.90 &0.10 &8.40 &0.10\\
N  & 8.36 &0.10 & 8.83 &0.20 &8.43 &0.20\\
O  & 9.02 &0.10 & 9.19 &0.12 &8.79 &0.06\\
S  & 7.57 &0.10 &\multicolumn{2}{c}{\nodata} &7.08 &0.06\\
Ne &\multicolumn{2}{c}{\nodata} & 8.67 &0.10 &8.25 &0.06\\
\bottomrule
\tabnotetext{a}{In~units~of~12~+~Log~$N$(X)/$N$(H).}
\end{tabular}
\end{table}

\subsection{Comparison of the oxygen abundances of the Orion nebula with those 
of B stars of the solar vicinity}

Przybilla, et al. (2008) based on the study of B stars of the solar vicinity have found that
12 + log O/H = 8.76 $\pm~0.03$, while Sim\'on-D\'{\i}az (2009 private communication) finds
that 12 + log O/H = 8.76 $\pm~0.04$ for the Orion association B stars, these results are in excellent 
agreement with that derived by Esteban et al. (2004) based on the O recombination lines, that 
amounts to $8.73 \pm~0.03$. The result by Esteban et al. (2004) includes a correction 
of 0.08 dex due to the fraction of O tied up in dust grains estimated by Esteban et al. (1998). 
By adopting the O dust correction of $ 0.12 \pm~0.03$~dex  estimated by Mesa-Delgado et al. (2009), 
the Orion nebula O abundance derived by Esteban et al. (2004) becomes $8.77 \pm~0.04 $, also in 
excellent agreement with those derived from the B stars.

On the other hand, based on the $T$(4363) method with $t^2$ = 0.00
Deharveng et al. (2000), Pilyugin, Ferrini, \& Shkvarun (2003), and Esteban 
et al. (2004) obtain for the Orion nebula 12 + log O/H values of 8.51, 8.49, 
and 8.51 respectively, values that after adding the fraction of dust tied up in dust grains 
are still smaller than those derived from B stars.

\subsection{Comparison of the oxygen abundances of the ISM of the solar vicinity with those of the Sun
and F and G stars of the solar vicinity}

In addition to the evidence presented in section 2 in favor
of large $t^2$ values, and consequently in favor of the \ion{O}{2}$_{RL}$ 
method, there is another independent test that can be used to discriminate 
between the $T$(4363) method and the \ion{O}{2}$_{RL}$ method that consists
in the comparison of stellar and  \ion{H}{2} region abundances of the 
solar vicinity. 

Esteban et al. (2005) determined that the gaseous O/H value derived from  \ion{H}{2} regions of the solar vicinity 
amounts to 12 + log (O/H) = 8.69, and including the fraction of O atoms tied up in dust grains it is
obtained that 12 + log (O/H) = 8.81 $\pm~0.04$ for the O/H value of the ISM of the solar vicinity. 
Alternatively from the 
protosolar value by Asplund et al. (2009), 
that amounts to 12 + log(O/H) = 8.71, and taking into account
the increase of the O/H ratio due to galactic chemical evolution since
the Sun was formed, that according to the chemical
evolution model of the Galaxy by Carigi et al. (2005) amounts to 0.13~dex, 
we obtain an O/H value of 8.84 $\pm~0.04$~dex, in very good agreement 
with the value based on the \ion{O}{2}$_{RL}$ method. In this comparison 
we are assuming that the solar abundances are representative of the abundances 
of the solar vicinity ISM when the Sun was formed.

\begin{table*}[!t]\centering
\setlength{\tabnotewidth}{1.495\columnwidth}
  \tablecols{6}
\setlength{\tabcolsep}{0.5\tabcolsep}
\caption{Primordial helium abundance}
\label{tta:disk}
\begin{tabular}{l@{\hspace{35pt}}c@{}r@{$\pm$}l@{}c@{\hspace{35pt}}c}
    \toprule
Method &&
\multicolumn{2}{c}{$Y_p$}&&
source\tablenotemark{a}\\
\midrule
\ion{H}{2}, ($t^2\ne0.000$)                && 0.2477 & 0.0029 && 1,2 \\
\ion{H}{2}, ($t^2 = 0.000$)                && 0.2523 & 0.0027 && 1,2 \\
WMAP + SBBN, ($\tau_n$ = $885.7 \pm 0.8$ s) && 0.2487 & 0.0006 && 3,4 \\
WMAP + SBBN, ($\tau_n$ = $881.9 \pm 1.6$ s) && 0.2479 & 0.0007 && 3,4,5,6,7 \\
WMAP + SBBN, ($\tau_n$ = $878.5 \pm 0.8$ s) && 0.2470 & 0.0006 && 3,5,7 \\
\bottomrule
\tabnotetext{a}{(1) Peimbert et al. (2007a); (2) Porter et al. (2005, 2007, 2009); (3) Dunkley et
al. (2009); (4) Arzumanov et al. (2000); (5) Serebrov et al. (2005, 2008); (6) Mathews et al. (2005);
(7) Peimbert (2008).}
\end{tabular}
\end{table*}

There are two other determinations of the present
O/H value in the ISM that can be made from 
observations of F and G stars of the solar vicinity. According to 
Allende-Prieto et al. (2004) the Sun appears deficient by roughly 
0.1 dex in O, Si, Ca, Sc, Ti, Y, Ce, Nd, and Eu, compared with its 
immediate neighbors with similar iron abundances, by adding this 
0.1 dex difference to the 
solar value by Asplund et al. (2009) we obtain a lower limit of 
12 + log O/H = 8.81 for the local interstellar medium.
A similar result is obtained from the data
by Bensby \& Feltzing (2006) who obtain for the six most O-rich 
thin-disk F and G dwarfs of the solar vicinity an average  
[O/H] = 0.16; by adopting their value as representative of the 
present day ISM of the solar vicinity we find 12 + log O/H = 8.87. 
Both results are in very good agreement with the O/H value derived 
from the \ion{O}{2}$_{RL}$ method.

\subsection{Comparison of the heavy element and helium abundances of M17 with those of K dwarf stars
and with models of galactic chemical evolution}

The best Galactic \ion{H}{2} region to determine the He/H ratio is M17 because it
contains a very small fraction of neutral helium and the error introduced by correcting
for its presence is very small. Carigi \& Peimbert (2008) obtained for M17 a value of 
$\Delta Y/\Delta Z = 1.97 \pm 0.41$ for $t^2 = 0.036 \pm 0.013$, where $Y$ and $Z$ are the helium and
heavy elements by unit mass and $t^2$ was determined
observationally. By correcting this value considering that the fraction of O trapped in dust amounts to 0.12 dex instead of 0.08 dex (Mesa-Delgado et al. 2009) we obtain $\Delta Y/\Delta Z~=~1.77~\pm~0.37$.
This $\Delta Y/\Delta Z$ value is in very good agreement with three independent
determinations: the one derived from the chemical evolution of the Galaxy for the galactocentric
distance of M17 that amounts to $\Delta Y/\Delta Z~=~1.70~\pm~0.4$, and two $\Delta Y/\Delta Z$ 
determinations derived from K dwarf stars of the solar vicinity that amount to $2.1~\pm~0.4$~(Jim\'enez et al. 2003) and to 
$2.1~\pm~0.9$  (Casagrande et al. 2007). On the other hand 
the value $\Delta Y/\Delta Z ~=~3.60~\pm~0.68$ derived from collisionally excited lines of M17 under
the assumption of $t^2 = 0.00$ is not in agreement with the chemical evolution models by Carigi \& Peimbert (2008)
nor with the values derived from K dwarf stars of the solar neighborhood.

\subsection{Comparison of the primordial helium abundance derived from \ion{H}{2} regions
together with that derived from WMAP and SBBN}

In Table 3 we present the determination of the primordial helium abundance, $Y_p$, based on
two different methods: a) that based on determination of  He/H ratios of metal poor \ion{H}{2} regions
and their extrapolation to the He/H value for the case of no heavy elements, and b) that based on the barion 
to photon ratio derived from WMAP observations together with the assumption of Standard Big Bang 
Nucleosynthesis, SBBN. Note that these two values might be different if SBBN does not apply to the primordial
nucleosynthesis stage of the universe (e.g. different number of neutrino families, varying gravitacional
constant, etc.).

In Table 3 we present two values derived from \ion{H}{2} regions (Peimbert, Luridiana
\& Peimbert 2007a): one for constant temperature inside the \ion{H}{2} regions, that based on the $T$(4363) value
($t^2 = 0.00$), and the other based on the assumption that the temperature varies over the observed volume
($t^2\ne0.00$). The adopted atomic physics for the helium recombination lines is that presented by
Porter et al. (2005, 2007, 2009). We also present three determinations based on the WMAP observations by
Dunkley et al. (2009), and three neutron lifetimes, $\tau_n$, those by: a) Arzumanov et al. 2000,
that amounts to $885.7 \pm 0.8 $ s, b) Mathews et al. (2005) that amounts to $881.9 \pm 1.6 $ s, and 
is based mainly on the average of the results by Arzumanov et al. (2000) and Serebrov et al. (2005), and c)
Serebrov et al. (2005, 2008) that amounts to  $878.5 \pm  0.8$ s. From Table 3 it is clear that the main source
of error in the WMAP + SBBN method to determine $Y_p$ is due to the neutron lifetime.

Moreover it follows from Table 3 that the $Y_p$ value for ($t^2\ne0.00$) is closer to the $Y_p$ value 
based on WMAP and SBBN than the $Y_p$ value for ($t^2 = 0.00$). Additional discussion on $Y_p$ is 
presented by Peimbert (2008) and by Peimbert et al. (2010).

\section{The ionization correction factor}
\label{sec:Ion}

To derive the total gaseous abundances of a given element it is necessary to
observe the relative amounts of all the ionization stages of that element 
present in a given nebula. Very often this is not possible and one has to correct for
the presence of the unobserved ions. This correction can be done by three different methods:
a) an empirical ionization correction curve, b) the use of equations, where the correction for the unseen ions
is estimated  by an ionization correction factor based on ratios
of other observed ions, and c) from tailor made photoionization models. 

Bowen and Wyse (1939) were the first to
propose the use of what has been called the empirical ionization distribution curve (e.g. Seaton 1960,
Aller, 1961, Aller and Liller 1968). While PC together with other authors started to use equations
with ionization correction factors (e. g. Seaton 1968, Rubin 1969, Peimbert and Torres-Peimbert 1971).

The calibration of the empirical ionization correction factors, $ICF$'s, has been obtained
by using photoionization models and by observing a given object at different lines of sight with
different degrees of ionization under the assumption of chemical homogeneity.

The ionization correction factors of N and Ne presented by PC have been used widely and are given by:

\begin{equation}
\frac{N(\rm N)}{N(\rm H)} = ICF({\rm N}) \frac{N({\rm N^+})}{N({\rm H^+})} = 
              \frac{N({\rm O})}{N({\rm O^+})}
              \frac{N({\rm N^+})}{N({\rm H^+})},
\label{eN1}
\end{equation}
and
\begin{eqnarray}
\nonumber
\frac{N({\rm Ne})}{N({\rm H})} & = &
             ICF({\rm Ne})       
             \left( \frac{N({\rm Ne}^{++})}{N({\rm H}^+)} \right) \\
                               & = & 
             \left( \frac{N({\rm O})}{N({\rm O}^{++})} \right)
             \left( \frac{N({\rm Ne}^{++})}{N({\rm H}^+)} \right),
\end{eqnarray}
where $N({\rm O}) =  N({\rm O^+}) + N({\rm O}^{++})$.

These two equations present  very good approximations to the total abundance ratios for those
cases where the $ICF$ is small, for large values of the $ICF$ they should be taken with caution.

For example in Figure 3 of Peimbert, Torres-Peimbert and Luridiana (1995) it can be seen that Ne$^{++}$/O$^{++}$
in planetary nebulae increases with decreasing density, indicating that for objects with
$N_e \leq 1000~{\rm cm}^{-3}$ equation (8) is a poor approximation to the 
Ne/O value; this result
probably is due to the presence of the charge exchange reaction

$$
{\rm O}^{+2} + {\rm H}^0 \rightarrow {\rm O}^+ + {\rm H}^+
$$

\noindent that permits the coexistence of Ne$^{++}$ with O$^+$ (e.g. 
Hawley \& Miller 1977, 1978; Hawley 1978; Pequignot, Aldrovandi \& 
Stasi\'nska 1978; Butler, Bender \& Dalgarno 1979; Pequignot 1980). Ionization structure
models predict that the lower the density the higher the H$^0$/H$^+$ ratio,
in agreement with the charge exchange suggestion and Figure 3.

Moreover for PNe and H II regions of low degree of ionization, those with a substantial fraction of
once ionized O, equation 4 only presents a lower limit to the total Ne/H abundance
(e. g. Figure 4 of Torres-Peimbert and Peimbert, 1977; and Figure 9 of Peimbert, Torres-Peimbert, \& Ruiz, 1992). 

Sets of useful ionization correction factor equations for gaseous nebulae have been presented by many authors
(e. g. Kingsburgh and Barlow 1994, Izotov et al. 2006).

\acknowledgments

\end{document}